\documentclass[a4paper,10pt,twocolumn,BCOR=4mm,DIV=calc]{scrartcl}
\usepackage{fixltx2e}[2006/09/13]  % package fixing some LaTeX bugs
\usepackage[american]{babel}
\usepackage[utf8]{inputenc}
\usepackage[margin=10pt,font=small,labelfont=bf,textfont=it,labelsep=endash, format=plain]{caption}
\usepackage[version=3]{mhchem} % Formula subscripts using \ce{}
\usepackage[dvips]{graphicx}   % final mode
\usepackage{subfig}
\usepackage{commath}  % Definition of derivatives
\usepackage{balance}
\usepackage{overcite}

\usepackage{charter}
\usepackage[bitstream-charter]{mathdesign}
\usepackage[protrusion=true, expansion=true, kerning=true,
 spacing=true,tracking=true,auto=true]{microtype} 
\microtypecontext{spacing=nonfrench}

\usepackage[breaklinks]{hyperref}
\usepackage{breakurl}
\hypersetup{%
  pdftitle = {Autocatalysis},
  pdfauthor = {R. Plasson, A. Brandenburg, L. Jullien, H. Bersini} ,
  pdfcreator = {\LaTeX\ with package \flqq hyperref\frqq},
  colorlinks=false,
  pdfborder=0 0 0 
}

\usepackage{indentfirst}

\renewcommand{\thefootnote}{\alph{footnote}}
\makeatletter
\renewcommand{\@makefnmark}{\textsuperscript{\texttt \thefootnote}}
\makeatother

\renewenvironment{abstract}{\centerline{\bf
Abstract}\vspace{0.5ex}\begin{quote}\small}{\par\end{quote}\vskip 1ex}

\usepackage{cleveref}

\title{Autocatalyses}

\author{Raphaël Plasson\thanks{Keio University, Yokohama, Japan}, Axel  Brandenburg\thanks{Nordita,
    Stockholm University, Stockholm, Sweden}, Ludovic Jullien\thanks{ENS, Paris, France}, Hugues
  Bersini\thanks{Iridia, ULB, Brussels, Belgium}}

\date{}

\KOMAoptions{DIV=last}

\begin{document}

\maketitle

\begin{abstract}
  Autocatalysis is a fundamental concept, used in a wide range of domains.  From the most general
  definition of autocatalysis, that is a process in which a chemical compound is able to catalyze
  its own formation, several different systems can be described. We detail the different categories
  of autocatalyses, and compare them on the basis of their mechanistic, kinetic, and dynamic
  properties. It is shown how autocatalytic patterns can be generated by different systems of
  chemical reactions.  The notion of autocatalysis covering a large variety of mechanistic
  realisations with very similar behaviors, it is proposed that the key signature of autocatalysis
  is its kinetic pattern expressed in a mathematical form.

\end{abstract}

\tableofcontents

\section{Introduction}

%%%% Historic %%%%%
The notion of ``autocatalysis'' was introduced by Ostwald in 1890\cite{Ostwald-90}
for describing reactions showing a rate acceleration as a function of time\cite{Ostwald-90}. It is for example the
case of esters hydrolysis, that is at the same time acid catalyzed and producing an organic acid
\cite{Laidler-86}. Defined as a chemical reaction that is catalyzed by its own products, it has
quickly been described on the basis of a characteristic differential equation
\cite{Ostwald-10*b,Ostwald-12}. Typically used to describe complex behaviors of chemical systems,
like oscillatory patterns \cite{Lotka-10}, it has immediately appeared to be essential for the
description of biological systems: growth of individual living beings \cite{Robertson-08},
population evolution \cite{Lotka-20} or gene evolution \cite{Muller-22}.

Extending this concept from a chemical description to a more open
context was initially carefully described as an analogy, sometime
qualified by the more general notion of ``autocatakinesis''
\cite{Lotka-25,Witzemann-33}. However, this eventually leads to an
overgeneralization of the term of autocatalysis, tending to be
assimilated to the notion of ``positive feedback'', for example in
economy \cite{Malcai.Biham.ea-02}.
%%% Article %%%

The notion of autocatalysis is now actively being used for describing
self-organizing systems, namely in the field of emergence of life and
artificial life. Autocatalytic processes are the core of the
mechanisms leading to the symmetry breaking of chemical compounds
towards homochirality \cite{Frank-53,Plasson.Kondepudi.ea-07}, and
could be identified in several experimental systems
\cite{Kondepudi.Kaufman.ea-90, Soai.Shibata.ea-95}. However, how such
autocatalytic processes shall manifest is still under heavy debate
\cite{Plasson-08, Blackmond-09}.

The purpose of this article is thus to clarify the
meaning of chemical autocatalysis and this effort will be undertaken
by covering these following points:
\begin{itemize}
\item What is autocatalysis for a chemical system? On the basis of the
  general description of a process allowing a chemical compound to
  enhance the rate of its own formation, autocatalysis is defined by a
  kinetic signature, expressed in a mathematical form.
\item How can an autocatalytic process be realized? As many mechanisms
  can reduce to the same macroscopic kinetic laws exhibiting
  autocatalysis, the focus is put on several mechanistic realisations
  of autocatalytic processes, based on simple models further
  illustrated by concrete chemical examples.
\item How can autocatalysis be observed and characterized? The
  focus is put on the dynamic properties, showing that this
  observable is the direct consequence of the kinetic pattern, rather
  than the underlying mechanism.
\item What is the role of autocatalysis? Embedded in non-equilibrium
  reaction network, the competition between autocatalytic processes
  allows the onset of chemical selection, that is the existence of bifurcation
  phenomena allowing the extinction of some compounds in favor of
  others.
\end{itemize}

\begin{figure*}[htb]
  \centering
    \includegraphics[width=17cm]{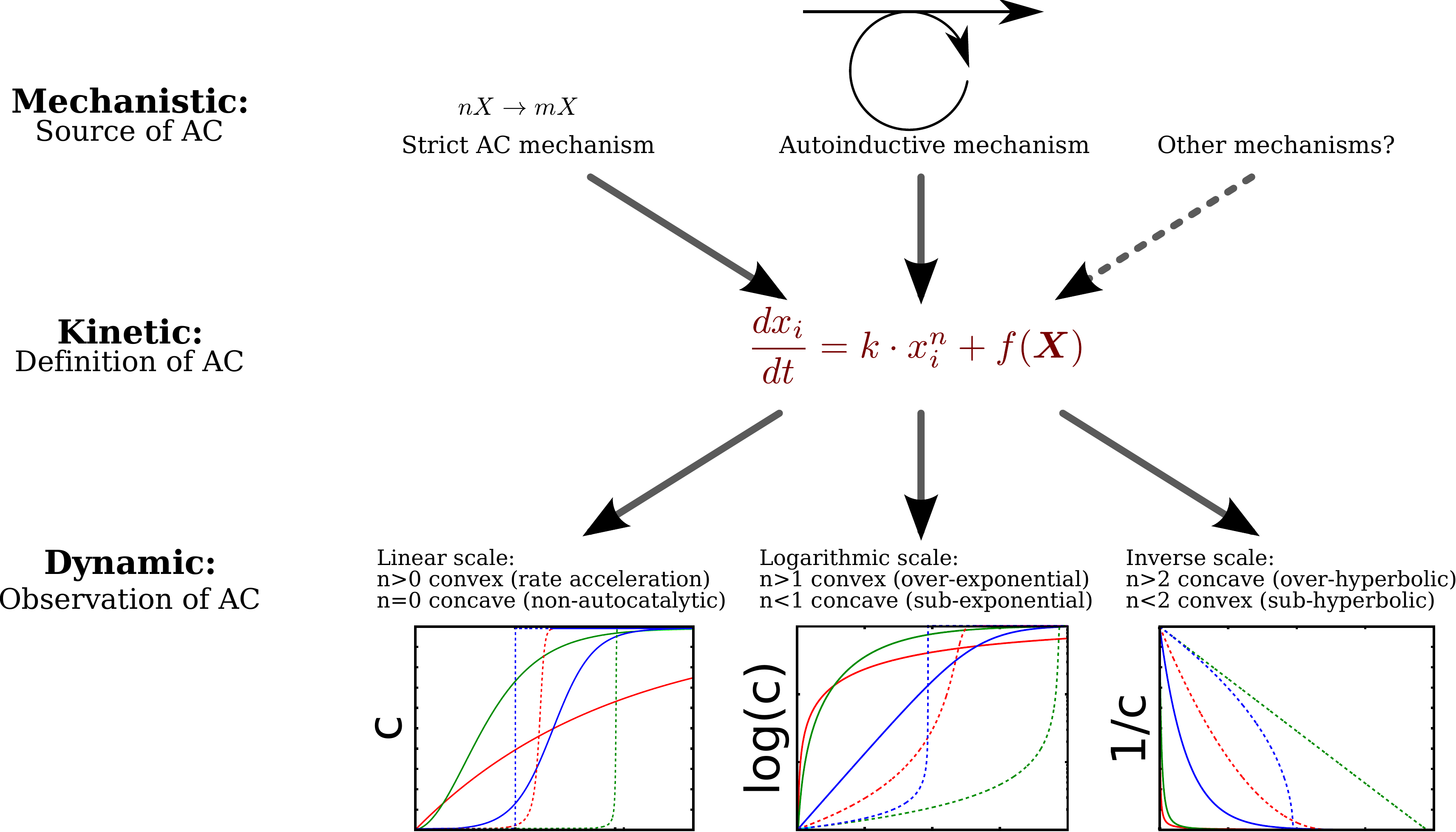}
    \caption{Classification of the concepts of autocatalysis (AC)
      depending on their descriptions (mechanistic, kinetic, and
      dynamic). The graphs represents the time evolution of a
      non-autocatalytic reaction ($n=0$, red), and of autocatalytic reactions
      of order $n=1/2$ (green), $1$ (blue), $3/2$ (dotted red), $2$
      (dotted green), and $3$ (dotted blue).}
    \label{fig0}
\end{figure*}

\section{Autocatalysis: a Practical Definition}
\label{sec:general-definition}

\subsection{A Kinetic Signature}
\label{sec:kinetic-concept}

\begin{figure}[hbt]
  \centering
    \includegraphics[width=8.25cm]{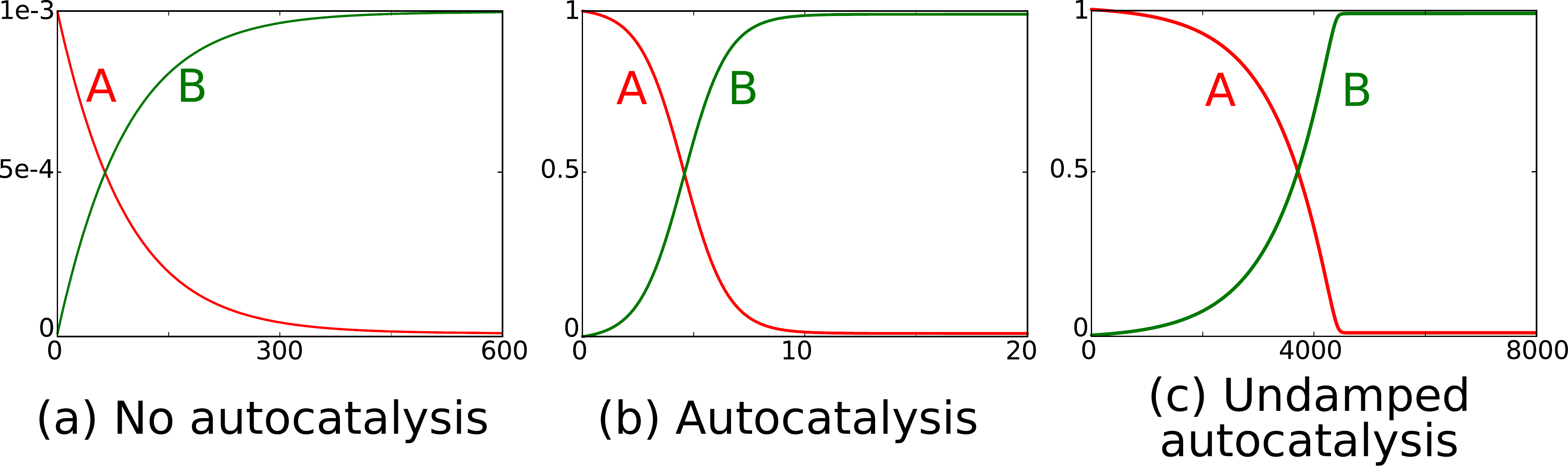}%
  \caption{(a-b): First order autocatalytic process ($\Gamma_1 =
      10^2$ M.s$^{-1}$) in presence of a non-autocatalytic reaction
      ($\Gamma_2 = 10^{-2}$ M.s$^{-1}$) of spontaneous transformation
      of A into B ($K_A = 1$ M, $K_B = 10^2$ M). (a) Diluted
      ($a_o=10^{-3}$~M). (b) Concentrated ($a_0=1$~M).  (c) Undamped
      autocatalysis (Indirect autocatalysis, described in \cref{fig2}(b),
      $\Gamma_4=0.1$ M.s$^{-1}$)}
  \label{fig:ACvsNAC}
\end{figure}

%%% Definition %%%

From its origin, the notion of autocatalysis has focused on the
kinetic pattern of the chemical evolution
\cite{Ostwald-10*b}. The general definition of
autocatalysis as a chemical process in which one of the products
catalyzes its own formation can be mathematically generalized as:
\begin{equation}
  \od{x_i}{t}=k(\mathbf{X})\cdot x_i^n+f(\mathbf{X}), \quad k>0 ;~ n>0;~|k|\gg|f|   \label{eq:general}
\end{equation}
$\mathbf{X}$ is the vector of all the concentrations $x_j$.  An
autocatalysis for the compound $x_i$ exists when the conditions of
\cref{eq:general} are fulfilled. The term
$k(\mathbf{X})\cdot x_i^n$ describes the autocatalytic process
itself, while $f(\mathbf{X})$ describes the sum of all other
contributions coming from the rest of the chemical system.

We have an effective practical definition of the concept of
autocatalysis, based on a precise mathematical formulation. The causes
of this kinetic signature can be investigated, searching what
mechanism is responsible for the autocatalytic term. This leads to the
discovery of a series of different kinds of autocatalysis processes,
and their respective effect, describing what observable behavior is
generated by the autocatalytic term (see \cref{fig0}).

\subsection{Potential vs Effective Autocatalysis}
\label{sec:effect-vs-struct}

This kinetic definition is purely structural. As a matter of fact, a
system may contain \emph{potential} autocatalysis i.e.\ an
autocatalytic core exists in the reaction network. However, in the
absence of some specific conditions necessary for this autocatalysis
to be \emph{effective}, the potential autocatalysis may be hidden by
other kinetic effects, and thus not manifests its behavior in
practice.

Possibly, in \cref{eq:general}, the term $f(\boldsymbol{X})$ may
simply overwhelm the autocatalytic process. This is typically the case
when an autocatalysis is present together with the non-catalyzed
version of the same reaction, that may not be negligible in all
conditions. A simple example is a system simultaneously
containing a direct autocatalysis \ce{$A$ + $B$ -> 2$B$}, concurrent with
the non autocatalytic reaction \ce{$A$ -> $B$}.  The autocatalytic
process follows a bimolecular kinetics, and will be more efficient in
a concentrated than in a diluted solution.  The dynamic profile of the
reaction is thus sigmoidal for high initial concentration of $A$, but
no more for low initial concentration (see \cref{fig:ACvsNAC}(a-b)).

It can also be seen that the term $k(\boldsymbol{X})$ may vary during
the reaction process. In a simple autocatalytic process as described
above, $k$ is proportional to the concentration in $A$, and is thus
more important at the beginning of the reaction (thus an initial
exponential increase of the product $B$) that at the end (thus a
damping of the autocatalysis) resulting in a global sigmoidal
evolution. In systems were the influence of $A$ on $k$ is weaker, as
detailed further, an undamped autocatalysis will be observed
characterized by an exponential variation until the very end (see
\cref{fig:ACvsNAC}(c)).

\section{Mechanistic Distinctions}
\label{sec:elementary-networks}

How can this kinetic pattern be realized? Let us now detail several
types of mechanisms. They can all be reduced, in some conditions, to
the autocatalysis kinetic pattern of \cref{eq:general}. All of
them will be equally defined in the paper as autocatalytic, while this
status may have been disputed in the past on account of the distinct
chemical realisations. In the following, we emphasize the major
mechanistic pattern to eventually be reduced to an equivalent kinetic
autocatalysis, and discuss where their difference comes from.

\subsection{Template Autocatalysis}
\label{sec:direct-autocatalysis}

% \subsubsection{Simple networks:}
% \label{sec:minimal-network-2}

\begin{figure}[bt]
  \begin{center}
    \subfloat[Direct]{\includegraphics[width=4cm]{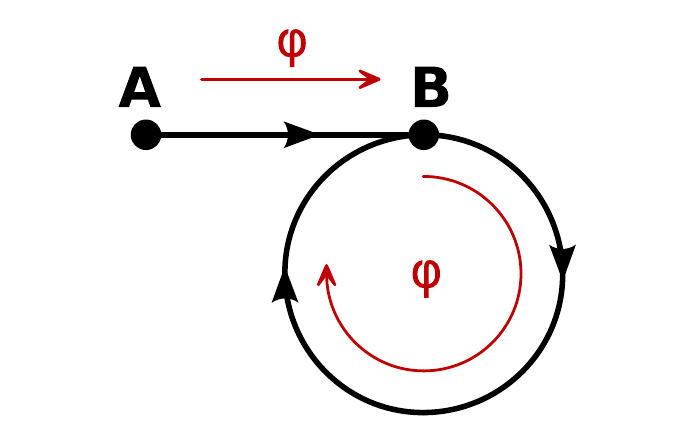}}\hspace{0.25cm}%
    \subfloat[Direct  with intermediate]{\includegraphics[width=4cm]{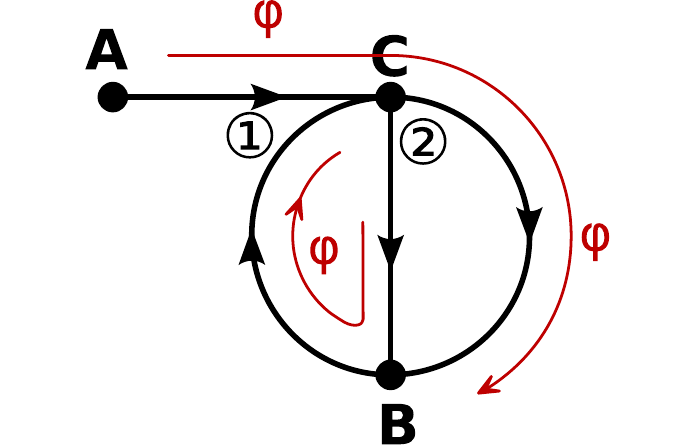}}
  
    \subfloat[Indirect]{\includegraphics[width=4cm]{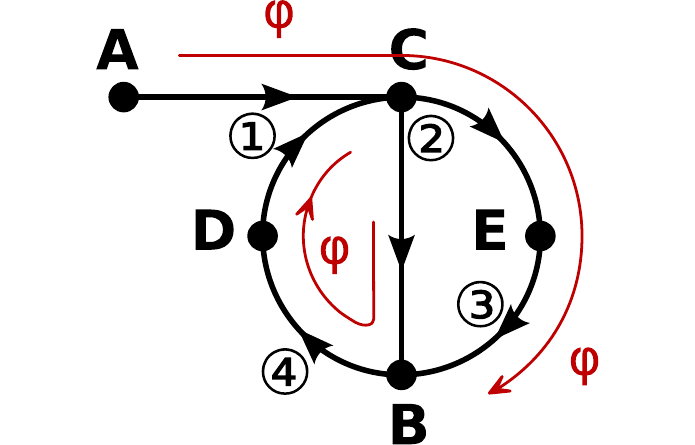}}\hspace{0.25cm}%
    \subfloat[Autoinductive]{\includegraphics[width=4cm]{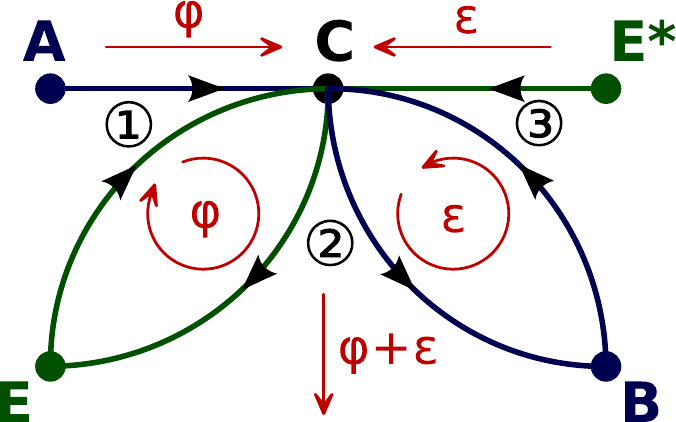}}
  
    \subfloat[Iwamura \emph{et al}\cite{Iwamura.Wells.ea-04} system]{\includegraphics[width=4.2cm]{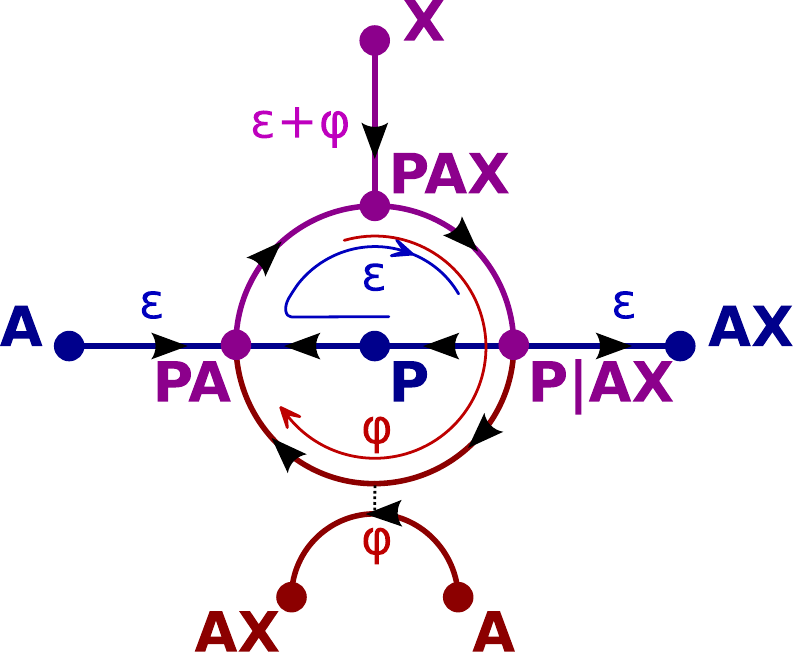}}\hspace{0.25cm}%
    \subfloat[Collective]{\includegraphics[width=3.8cm]{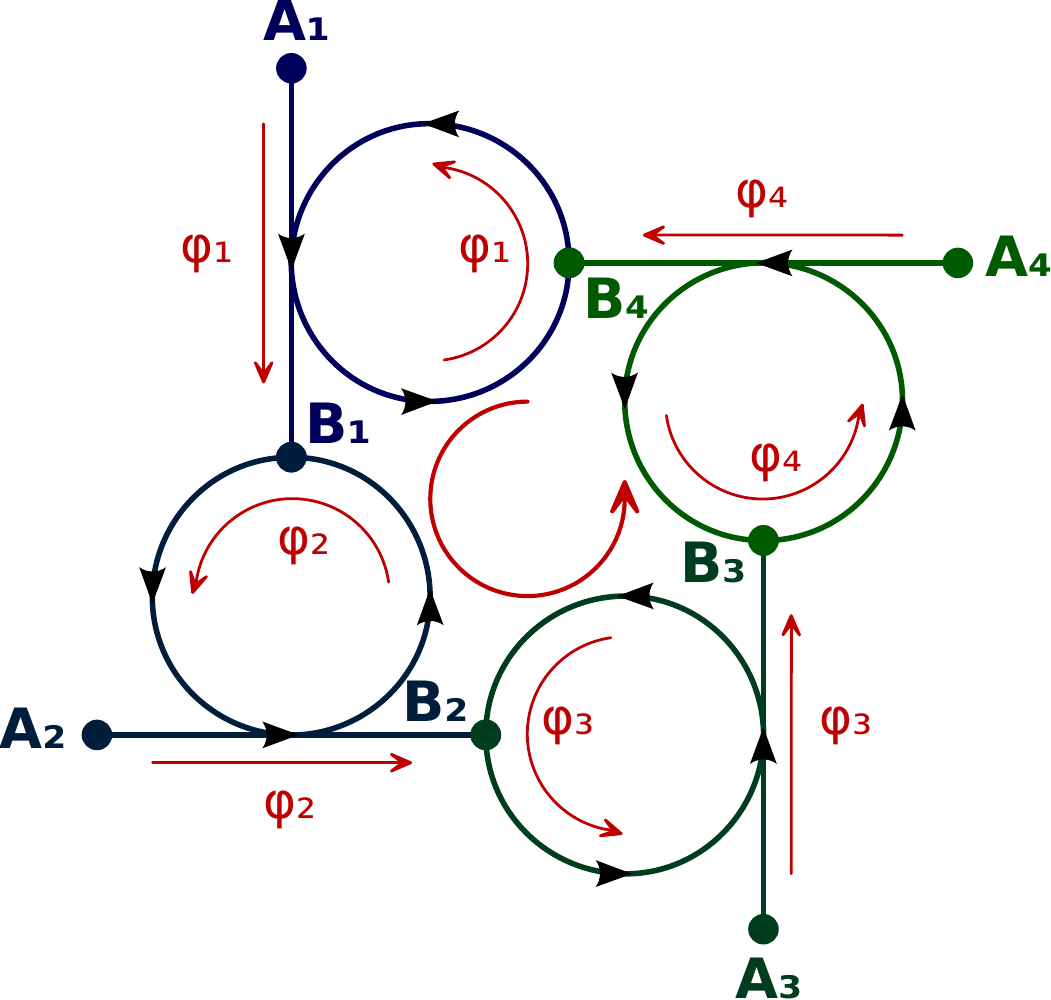}}
    \caption{Reaction network of different autocatalytic processes of
      spontaneous transformation of $A$ into $B$ (a-d), of $A+X$ into
      $AX$ (e), and of $A_i$ into $B_i$ (f). The indicated fluxes
      correspond to what is observed within the QSSA.}
    \label{fig1}
  \end{center}
\end{figure}

The simplest autocatalysis is obtained by the \ce{$X$ -> 2$X$}
pattern. It can be represented by:
\begin{equation}
  \ce{$A$ + $B$  <=>[k_1][k_{-1}]  $B$ + $B$}
\end{equation}
The corresponding network is given in \cref{fig1}(a).  It can
further be decomposed through the introduction of an intermediate
compound $C$:
\begin{align}
  \ce{$A$ + $B$}&\ce{<=>[\Gamma_1]  C} \\
  \ce{C}&\ce{  <=>[\Gamma_2]  $B$ + $B$}
\end{align}
The corresponding network is  given in \cref{fig1}(b).

The first mechanism entails the following kinetic evolution:
\begin{align}
  \dod{b}{t} &= - \dod{a}{t}\\
  &= k_1ab-k_{-1}b^2
\end{align}
This can be expressed as a chemical flux $\varphi=\od{b}{t}$, by relying on the
Mikulecky formalism
\cite{Peusner.Mikulecky.ea-85,Mikulecky-01,Plasson.Bersini-09}:
\begin{align}
  \varphi &= \Gamma_1(V_AV_B-V_B^2)\\
  V_A&=\frac{a}{K_A} \\
  V_B&=\frac{b}{K_B} \\
  \Gamma_1&=k_1\cdot K_AK_B= k_{-1} \cdot K_B^2
\end{align}
$k_1$ and $k_{-1}$ are the kinetic constant rates of the reaction $1$ in the direct and reverse
direction. $K_A$ and $K_B$ are the thermodynamic constant of formation of compounds $A$ and $B$. 

Formally there is a linear flux $\varphi$ of transformation of $A$
into $B$, coupled to a circular flux of same intensity from $B$ back
to $B$ (see \cref{fig1}(a-b)). In presence of an intermediate
compound, the equations becomes:
\begin{align}
  \varphi_1&= \Gamma_1(V_AV_B-V_C)\\
  \varphi_2&=\Gamma_2(V_C-V_B^2)
\end{align}

\begin{figure}[bt]
  \begin{center}
    % \subfloat[Direct autocatalysis]{\includegraphics[width=2in]{autocat_direct_logscale}}
    
    \subfloat[Direct]%
    {\includegraphics[width=4cm]{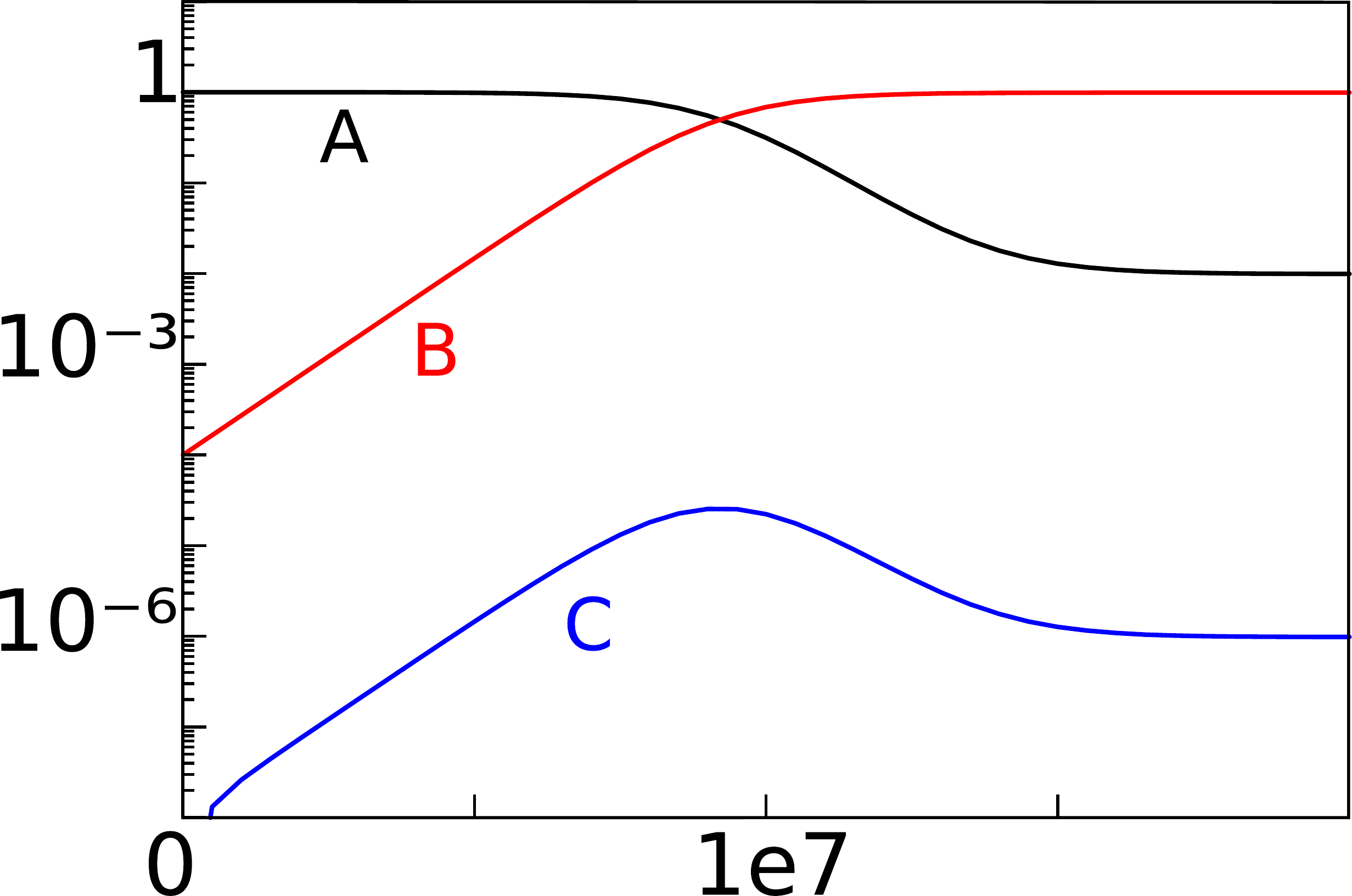}}\hspace{0.25cm}%
    \subfloat[Indirect]%
    {\includegraphics[width=4cm]{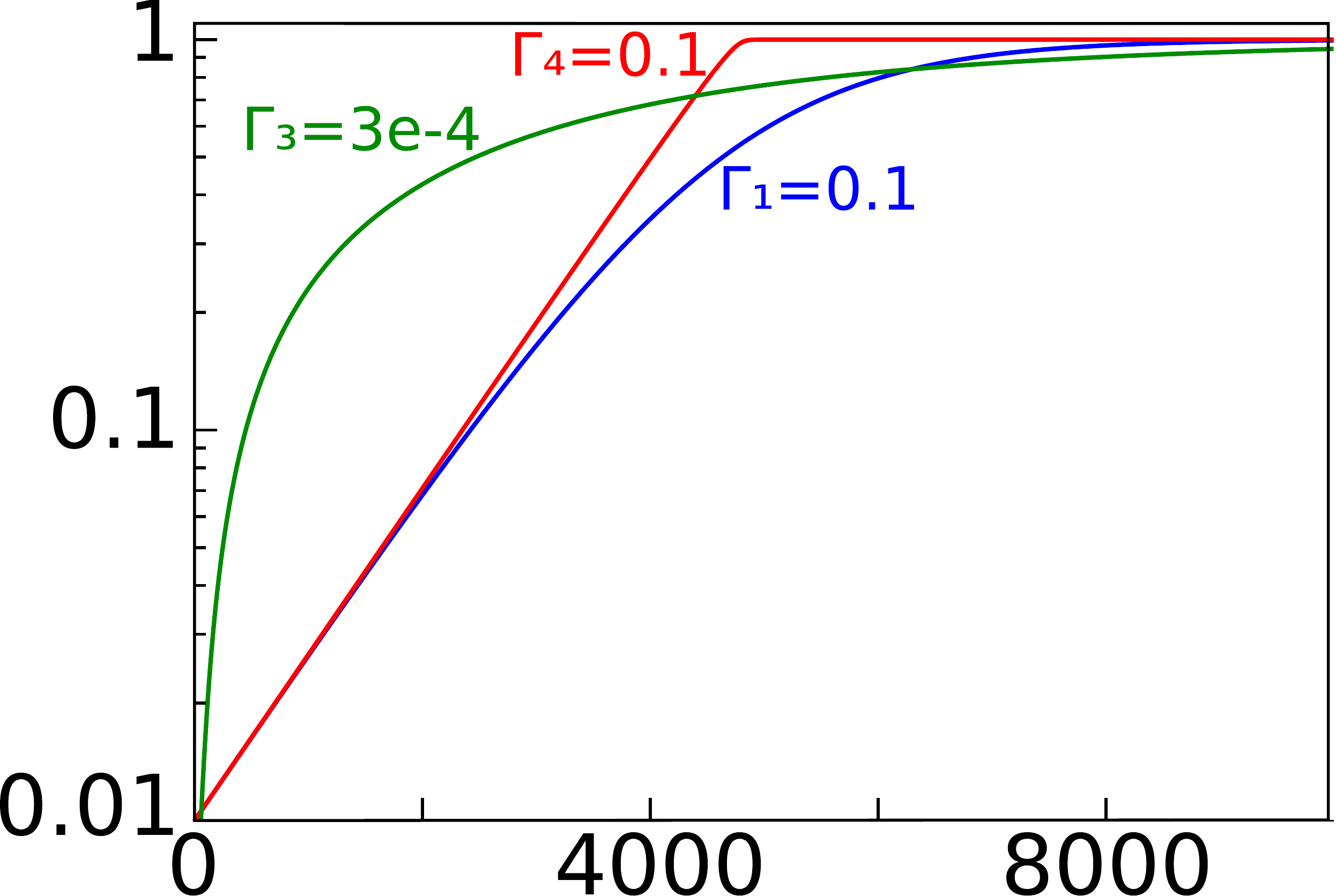}}
    
    \subfloat[Autoinductive]{\includegraphics[width=4cm]{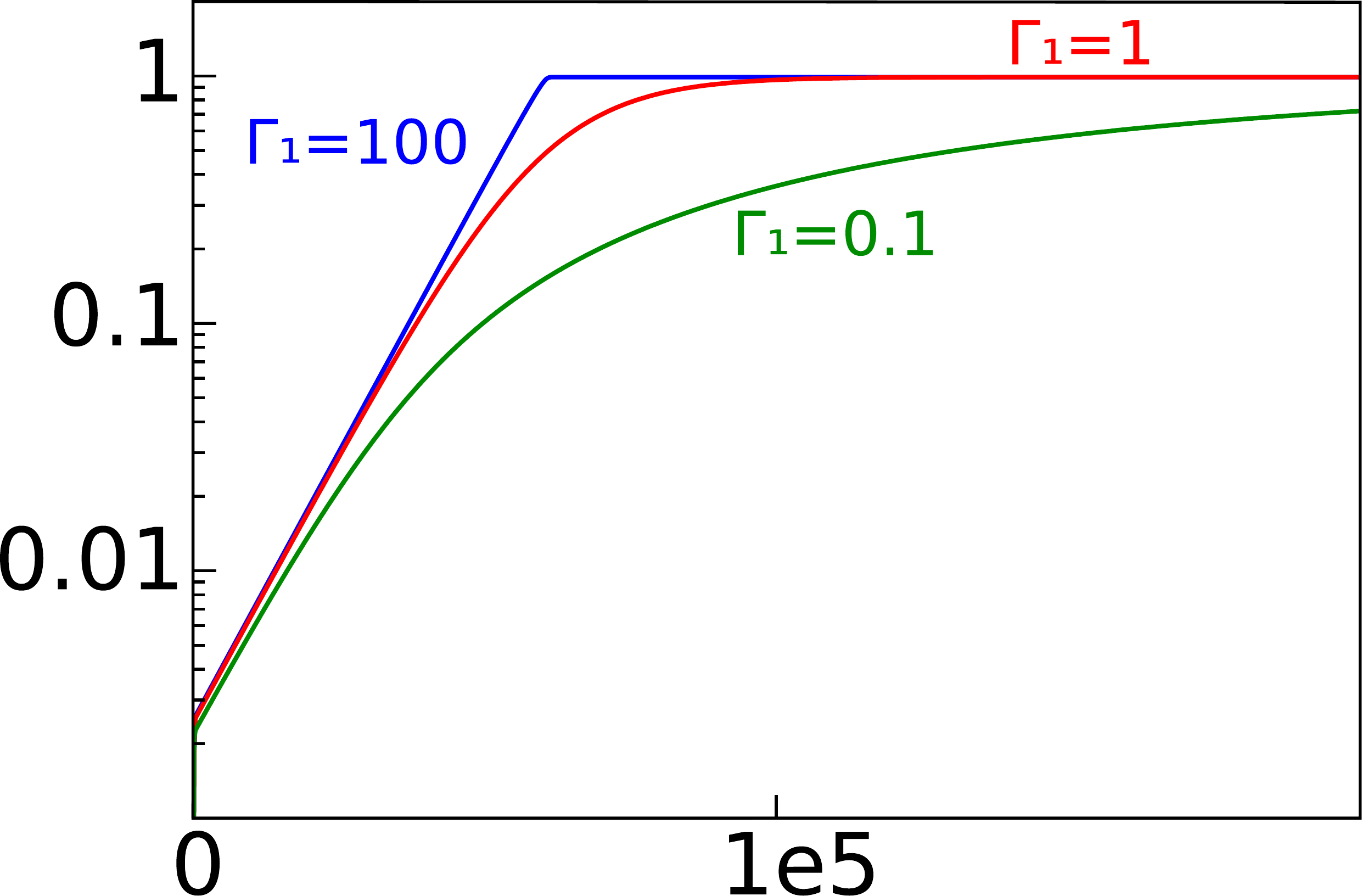}}\hspace{0.25cm}%
    \subfloat[Collective]{\includegraphics[width=4cm]{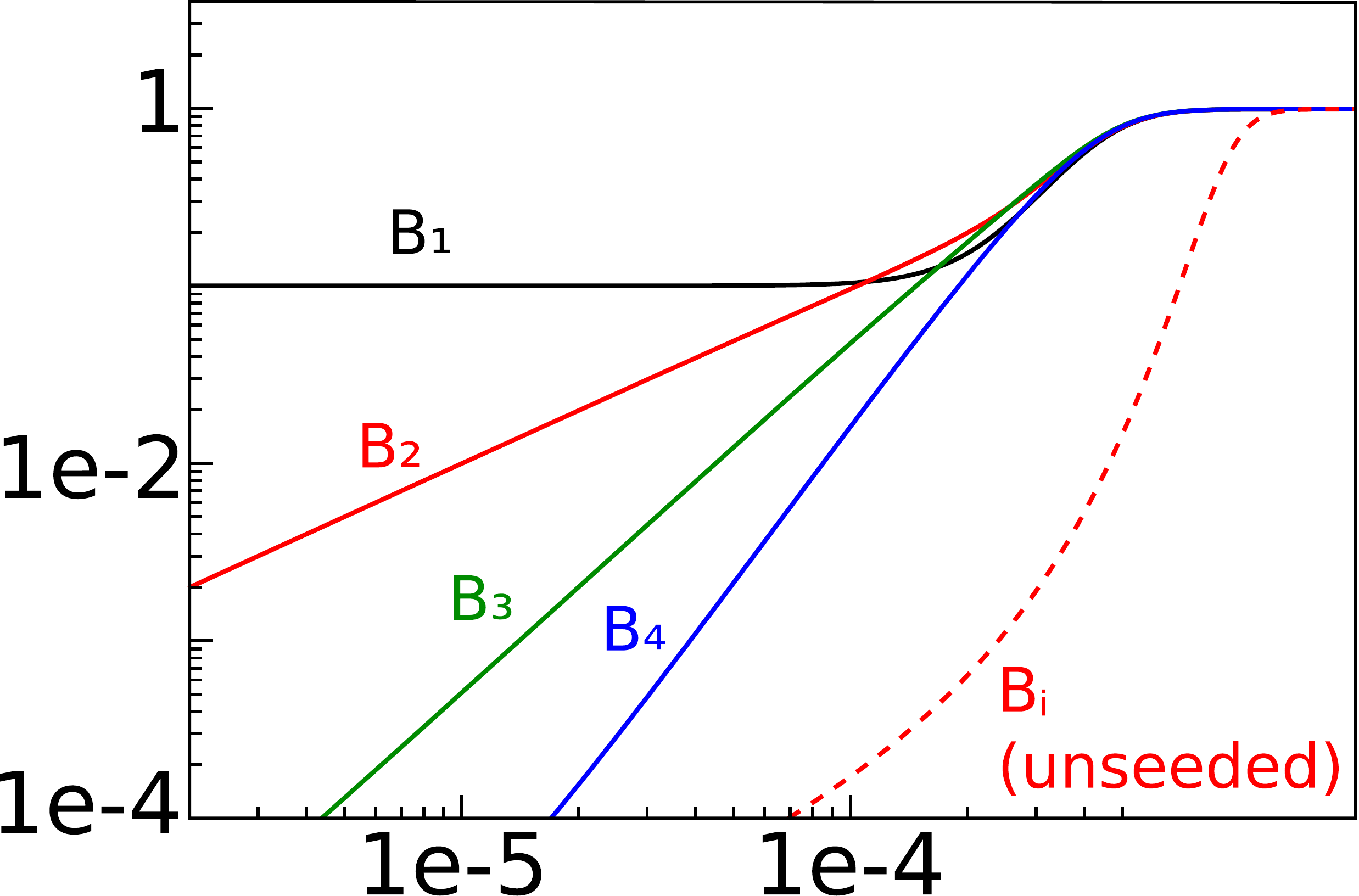}}
    \caption{Time evolution of compound concentrations for different
      autocatalytic processes of spontaneous transformation of $A$
      into $B$ ($K_A=1$~M and $K_B=100$~M) in a
      logarithmic scale for concentrations (a-c), or logarithmic
      scales for both time and concentrations (d).  $K$ and concentrations
       are in M, times in s, and $\Gamma$ in M.s$^{-1}$. (a):
      \cref{fig1}(b), $\Gamma_1=1$, $\Gamma_2=10^{-4}$, $K_C=0.01$;
      (b): \cref{fig1}(c), $\Gamma_1=\Gamma_2=\Gamma_3=\Gamma_4=10$
      (except the values indicated on the graph),
      $K_C=K_D=K_E=0.01$; (c): \cref{fig1}(d),
      $\Gamma_2=\Gamma_3=100$, $K_C=K_E=1$, $K_{E^*}=10$; (d):
      \cref{fig1}(f), $\Gamma_1=100$, $\Gamma_2=1$.}
  \label{fig2}
  \end{center}
\end{figure}

Under the hypothesis that $C$ is an unstable intermediate, (i.e.\ $K_C
\ll K_B , K_A$), 
the variation of $C$ can be neglected compared to the
variations of $A$ and $B$ (quasi steady-state approximation, hereafter
QSSA), so that:
\begin{align}
  \varphi_1&\simeq\varphi_2 \\
  &=\varphi\\
  \Rightarrow \quad \varphi&=
  \frac{\Gamma_1\Gamma_2}{\Gamma_1 + \Gamma_2}
  (V_AV_B-V_B^2)
\end{align}
The system is strictly equivalent to the direct autocatalysis, with an
apparent rate $ {\Gamma_1\Gamma_2}/(\Gamma_1 + \Gamma_2)$. With
these two systems, we are in presence of the perfect kinetic signature
of an autocatalytic system, following a sigmoidal evolution (see
\cref{fig2}(a)). This equivalence is guaranteed as long as the
compound $C$ remains unstable. When it is not the case, the dimeric
intermediate $C$ hardly liberates the final compound $B$, which
eventually leads to an autocatalytic process of order $1/2$ rather than $1$
\cite{Kiedrowski-93,Wills.Kauffman.ea-98}.

% \subsubsection{Example:}
% \label{sec:example-1}

Template autocatalysis requires a direct association between the reactants and the products. This is
typically the case of DNA replication, one double strand molecule giving birth to two identical
double strand molecules, thanks to the very selective association of complementary nucleotides along
each strand.  More simple examples can be found in some biological mechanisms that requires
autocatalytic processes, for example for the generation of chemical oscillation inducing circadian
rhytmicity in cells. The system described by Mehra \emph{et al}\cite{Mehra.Hong.ea-06} is based on a non
equilibrium system of association/dissociation of proteins forming a large chemical cycle [$C$
  \ce{->} $AC$ \ce{->} $AC^*$ \ce{->} $ABC^*$ \ce{->} $BC^*$ \ce{->} $C^*$ \ce{->} $C$], maintained by a flux of ATP
consumption, one cycle consuming and freeing $A$ and $B$\cite{Mehra.Hong.ea-06}. The oscillations
are generated by coupling this chemical flux to an autocatalytic process of phosphorylation obeying
to the reaction scheme\cite{Wang.Wu-02}: \ce{$A$ + $C$ + $AC^*$ -> 2$AC^*$}.

\subsection{Network Autocatalysis}
\label{sec:indir-autoc-2}

The direct mechanism of template autocatalysis is conceptually the
simplest framework. It may actually not be the most representative
class of autocatalysis, and a similar kinetic signature can result
from more complex reaction networks.

\subsubsection{Indirect Autocatalysis:}
\label{sec:minimal-network}

The autocatalytic effect can be indirect when reactant and products
never directly interact:
\begin{align}
  A+D & \ce{<=>[\Gamma_1]  $C$} \label{eq:3}\\
  C & \ce{<=>[\Gamma_2]} B + E \label{eq:4}\\
  E & \ce{<=>[\Gamma_3]}  B  \label{eq:5}\\
  B & \ce{<=>[\Gamma_4]}  D \label{eq:6}
\end{align}
There is no direct $A/B$ coupling, nor direct $2B$ formation, but the
presence of a dimeric compound $C$. The network decomposition of this
system (see \cref{fig1}(c)) implies once again a linear flux of
transformation of $A$ into $B$, linked to a large cycle of reaction
transforming $B$ back to $B$.  This system is still reducible to an
\ce{$X$ -> 2$X$} pattern.

The QSSA for compounds $C,D,E$ allows to express the reaction flux as:
\begin{align}
  \varphi &= \frac{1}{\frac{1}{\Gamma_1} + \frac{1}{\Gamma_2} + \frac{V_A}{\Gamma_4}  + \frac{V_B}{\Gamma_3}}
  \left(
    V_AV_B - V_B^2
  \right)
\end{align}
The details of the calculations are given in appendix.

When the terms $V_A/\Gamma_4$ and $V_B/\Gamma_3$ are small compared to either $\Gamma_1^{-1}$ or
$\Gamma_2^{-1}$ (i.e.\ when at least one of the two reactions of \cref{eq:3}-\cref{eq:4}) is
kinetically limiting), the system behaves like a simple autocatalytic system, 
with $\varphi \propto
a\cdot b$ before the reaction completion, with a progressive damping of the exponential growth as
long as $A$ is consumed. When the term $V_A/\Gamma_4$ is predominant (i.e\ when the reaction of
\cref{eq:6} is kinetically limiting), the flux is $\varphi \propto b$: the profile remains exponential up to the
reaction completion, with no damping due to $A$ consumption. When the term $V_B/\Gamma_3$ is
predominant (i.e\ when the reaction of \cref{eq:5} is kinetically limiting), the flux is $\varphi \propto a$:
the autocatalytic effect is lost (see \cref{fig2}(b)).

Network autocatalysis is probably the most common kind of
mechanisms. A typical biochemical example is the presence of
autocatalysis in glycolysis
\cite{Ashkenazi.Othmer-77,Nielsen.Sorensen.ea-97}. In this system,
there is a net balance following the \ce{$X$ -> 2$X$} pattern. ATP
must be consumed to initiate the degradation of glucose, but much more
molecules of ATP are produced during the whole process. While these
systems are effectively autocatalytic, there is obviously no possible
``templating'' effect of one molecule of ATP to generate another one.

\subsubsection{Collective Autocatalysis:}
\label{sec:coll-autoc}

More general systems, reminiscent of the Eigen's hypercycles
\cite{Eigen.Schuster-77}, are responsible of even more indirect
autocatalysis. No compound influences its own formation rate, but
rather influences the formation of other compounds, which in turn
influence other reactions, in such a way that the whole set of
compounds collectively catalyzes its own formation.

A simple framework can be built from the association of several
systems of transformation \ce{$A_i$ -> $B_i$}, each $B_i$ catalyzing
the next reaction (see \cref{fig1}(f)):
\begin{align}
  A_i+ B_{i-1} & \ce{<=>[\Gamma_i]}  B_i + B_{i-1}\\
  &(\mathrm{with} ~ i=\{1,2,3,4\} ~ \mathrm{and} ~ B_4 \equiv B_0) \nonumber
\end{align} 
There are four
independent systems, only connected by catalytic activities.

If the system is totally symmetric, then all $b_i$ are equal,
and all $a_i$ are equal, so that the rates become:
\begin{align}
  \varphi_i&=\Gamma_i V_{B_{i-1}}(V_{A_i}-V_{B_i})\\
  \varphi&=\Gamma V_B(V_A-V_B)
\end{align}
This leads to a \emph{collective} autocatalysis with all 
compounds present. They mutually favor their formation, which
results in an exponential growth of each compound (see
\cref{fig2}(d) dotted curve).

With symmetrical initial conditions (i.e.\ identical for the four
systems), the system strictly behaves autocatalytically. If the
symmetry is broken, e.g.\ by seeding only one of the $B_i$, the system
acts with delays. The evolution laws are sub-exponential, of
increasing order; at the very beginning of the reaction, considering
that $A_i$ do not significantly change and that $B_i$ are in low
concentrations, we obtain $\varphi_i \propto t^{i-1}$. Seeding with
$B_1$, the compound $B_2$ evolves in $t^2$. Its impact on compound $B_3$
induces an evolution in $t^3$. In its turn, the impact of compound $B_3$
on compound $B_4$ induces an evolution in $t^4$. The compound $1$ at
first remains constant, and it is only following a given delay that it
gets catalyzed by $B_4$ (see \cref{fig2}(d)).

This system is actually not characterized by a direct cyclic flux,
but by a cycle of fluxes influencing each other and resulting in a
cooperative collective effect:
\begin{align}
  \label{eq:2}
    (A_1 + A_2 + A_3 + A_4) + (B_1 + B_2 + B_3 + B_4) &\\
 \ce{->} \quad 2 (B_1 + B_2 + B_3 + B_4)&
\end{align}
The simultaneous presence of all different compounds is needed to
observe a first order autocatalytic effect. Given asymmetric initial
conditions, a transitory evolution of lower order is first observed,
until the formation of the full set of compounds.

A typical example of collective autocatalysis is observed for the
replication of viroids \cite{Flores.Delgado.ea-04}. Each opposite
strand of cyclic RNAs can catalyze the formation of the other one,
leading to the global growth of the viroid RNA in the infected cell.

\subsubsection{Template vs Network Autocatalysis:}
\label{sec:conclusion-1}

All the preceding systems can be reduced to a \ce{$X$ -> 2$X$}
pattern.  This is characterized by a linear flux of chemical
transformations, coupled to an internal loop flux: for each molecule
(or set of molecules) $A$ transformed into $B$, one $B$ is transformed
and goes back to $B$, following a more or less complex pathways. They
can be considered as mechanistically equivalent: a seemingly direct
autocatalysis may really be an indirect autocatalysis once its precise
mechanism is known, decomposing the global reaction into several
elementary reactions.

Practically, autocatalysis will be considered to be direct (or
template) when a dimeric complex of the product is formed
(i.e.\ allowing the ``imprint'' of the product onto the reactant).  If
such template complex is never formed, we preferentially speak of network
autocatalysis, in which the \ce{$X$ -> 2$X$} pattern only results from the
reaction balance.

\subsection{Autoinductive Autocatalysis}
\label{sec:netw-autoc}

Some reactions are not characterized by a \ce{$X$ -> 2$X$} pattern,
but still exhibit a mechanism for the enhancement of the reaction rate by
the products. This is typically the case for systems where the
products increase the reactivity of the reaction catalyst rather than
directly influencing their reaction production itself. These systems
still possess the kinetic signature of \cref{eq:general}, but are
sometime referred as ``autoinductive'' instead of
``autocatalytic'' \cite{Blackmond-09}. 

 \subsubsection{Simple network:}
 \label{sec:minimal-network-1}

Let us take a simple reaction network of a transformation \ce{$A$ -> $B$} catalyzed by a compound
that can exist under two forms $E/E^*$, $E^*$ being the more stable one. These two forms of the
catalyst interact differently with the product $B$ (see \cref{fig1}(d)):
\begin{align}
  A + E & \ce{<=>[\Gamma_1]}  C \\
  C     & \ce{<=>[\Gamma_2]}  B + E \\
  C     & \ce{<=>[\Gamma_3]}  B + E^*
\end{align}
There is no dimeric compound in the system, even indirectly formed.

Provided the catalyst, present in $C$, $E$, and $E^*$, is in low total concentration, the QSSA
implies the presence of two fluxes: the transformation of $A$ into $B$ catalyzed by $E$ of intensity
$\varphi$, and the transformation of $E^*$ into $E$ catalyzed by $B$ of intensity $\varepsilon$,
with $\varphi\gg \varepsilon$. Assuming that $E^*$ is very stable compared to $E$ and $C$, this
decomposition gives (see appendix for details):
\begin{align}
  \varphi &=
   \frac{ \Gamma_1\Gamma_2V_{E^*}^0}{\Gamma_1V_A +
     \Gamma_2 V_B} 
   (  V_BV_A - V_B^2)
\end{align}

The autoinduction is kinetically equivalent to the indirect
autocatalysis mechanism:
\begin{itemize}
\item When $\Gamma_2 \gg \Gamma_1\frac{K_B}{K_A}$, the flux $\varphi$ is
  $\Gamma_1V_{E^*}^0(V_A- V_B)$: the system is
  non-autocatalytic.
\item When $\Gamma_2 \approx \Gamma_1\frac{K_B}{K_A} $, the flux $\varphi$ is $\Gamma_2\frac{V_{E^*}^0}{V_A^0}(V_AV_B- V_B^2)$: the system is
  simply autocatalytic.
\item When $\Gamma_2 \ll \Gamma_1\frac{K_B}{K_A}$ , the flux $\varphi$ is
  $\Gamma_2V_{E^*}^0
  \left(
    V_B-\frac{V_B^2}{V_A}
  \right)$: the system presents an undamped
  autocatalysis.
\end{itemize}

Following the kinetic analysis, the behavior is similar to the time
evolution of autocatalytic systems (See \cref{fig2}(c)). The
behavioral equivalence of these two systems (kinetically equivalent
but mechanistically very different) will be investigated in more details in
the next section.

\subsubsection{Iwamura's model}
\label{sec:iwamuras-model-1}

An example of autoinductive autocatalysis is the proline-catalyzed $\alpha$-aminoxylation of
aldehydes\cite{Iwamura.Wells.ea-04}. The core principle is a reaction \ce{$A$ + $X$ -> $AX$},
catalyzed by $P$, the product $AX$ catalyzing the first catalytic step \ce{$P$ + $A$ -> $PA$} (see
\cref{fig1}(e)). This chemical system can be decomposed into two different fluxes \ce{$A$ + $X$ ->
  $AX$}, one coupled to a catalytic cycle [$P$  \ce{->}  $PA$  \ce{->}  $PAX$  \ce{->}  $P|AX$  \ce{->}  $P$], and one
coupled to a catalytic cycle [$PA$  \ce{->}  $PAX$  \ce{->}  $P|AX$   \ce{->}  $PA$]. The first one contains
the slow reaction of $A$ on $P$, and corresponds to a slow flux $\varepsilon$. The second one only
contains fast reactions, and corresponds to a fast flux $\varphi$. In an ideal case (see appendix
for details), the flux of production of $AX$ is equal to:
\begin{align}
  \label{eq:1}
  \varphi = \Gamma_5 V_P^0
  \left(
    V_AV_{AX}-\frac{V_{AX}^2}{V_X}
  \right)
\end{align}
The kinetic signature of an undamped autocatalysis is once again obtained.

\subsubsection{Network vs Autoinductive Autocatalysis:}
\label{sec:indir-vs-auto}

Autoinductive autocatalysis is mechanistically different from network
or template autocatalysis. The balance equation is rather of the form
\ce{$A$ + $\alpha$ $B$ -> $(1+\alpha)$ $B$}, with $\alpha \ll 1$. The linear
transformation \ce{$A$ -> $B$} is only weakly coupled to the cycle
of $B$ back to itself, this latter one being subject to a much lower
flux than the linear flux. However, autoinduction is kinetically and
dynamically equivalent to network autocatalysis, leading to the same
kind of differential equation, and thus of behavior.  It must be noted
that the undamped exponential profile---due to a flux only proportional
to the products and not to the reactant---is not characteristic of
autoinductive processes \cite{Iwamura.Wells.ea-04} but can also be
explained by network autocatalytic mechanisms, when the consumption of
the reactant is not limiting the kinetic of the network.

\section{Embedded Autocatalyses}
\label{sec:competitions}

Autocatalysis is not so important \emph{per se} but as a way of giving
birth to rich non-linear behaviors like bifurcation, multistability or
chemical oscillations. It is crucial to study the interaction of
autocatalytic mechanisms and their ability to generate such behaviors
when embedded in a larger chemical network.

\subsection{Dynamical Distinctions}
\label{sec:autocatalysis-order}

Different behaviors depending on the order $n$ of the autocatalysis
can be observed in biochemical competitive systems. They are
classically studied in population evolution \cite{Szathmary-91,Nowak-06} and
described as ``survival of the all'' in the case of $0<n<1$
(characterized by the coexistence of all compounds), as ``survival of
the fittest'' in the case of $n=1$ (when the only stable solution
retains the fittest compound or the most "reproductible") and as
``survival of the first'' in the case of $n>1$ (when the final
solution just retains the product initially present in the highest
concentration).

The case $0<n<1$ is the least interesting one, as it hardly leads to a
clear selectionnist process. However, real mechanism that seems to
possess a first order autocatalysis may actually present a lower
autocatalytic order. This is typically the case for direct template
autocatalysis, in which the order falls to $1/2$ on account of the
high stability of the dimeric intermediate---which is actually a
necessary condition for the selectivity of template replication
\cite{Kiedrowski-86,Kiedrowski-93,Wills.Kauffman.ea-98}. This turns
out to be a fundamental problem for understanding the emergence of the
first replicative molecules
\cite{Szathmary.Gladkih-89,Lifson.Lifson-99,Scheuring.Szathmary-01}.

More complex mechanisms may lead to higher orders,
typically by the formation of dimeric
autocatalysts \cite{Wagner.Ashkenasy-09*b}.  This is the case
of the Soai reaction whose high sensitivity to initial conditions
may potentially be explained by the formation of
trimeric \cite{Gridnev.Serafimov.ea-03} or even hexameric
complexes \cite{Schiaffino.Ercolani-08}.

\subsection{Comparative Efficiency of Direct and Autoinductive Autocatalyses}
\label{sec:comp-effic-direct}

The relative efficiency of two different autocatalytic mechanisms can be evaluated by having them
competing which each other. Bifurcations appear when these two autocatalytic processes are placed in
a nonequilibrium open-flow system, both being fed by the same incoming compound and with
cross-inhibition between them:
\begin{align}
            &  \ce{->}                           A&           &  \text{(incoming flux)}\\
  A         &  \ce{<=>[\alpha]}  B_1      &                   & \text{(Direct AC)} \label{eq:mech1}\\
  A         & \ce{<=>[\beta]}      B_2      &                 & \text{(Autoinduced  AC)} \label{eq:mech2}\\
  B_1 + B_2 &   \ce{->}                 (P)     &             & \text{(cross inhibition)}\\
  B_1       &   \ce{->}                            &          & \text{(outgoing flux)}\\
  B_2       &   \ce{->}                            &           & \text{(outgoing flux)}
\end{align}
In the case of total symmetry between $B_1$ and $B_2$, with the same direct autocatalystic
mechanism, this system would correspond to the classical Frank model for the emergence of
homochirality \cite{Frank-53}. Because of the system symmetry, the same probability to end up with
either $B_1$ or $B_2$ is observed.

\begin{figure}[bt]
  \begin{center}
    \subfloat[Sharp bifurcation depending on the relative values of
    $\alpha$ and $\beta$ for moderate
    reactivities.]{\includegraphics[width=8cm]{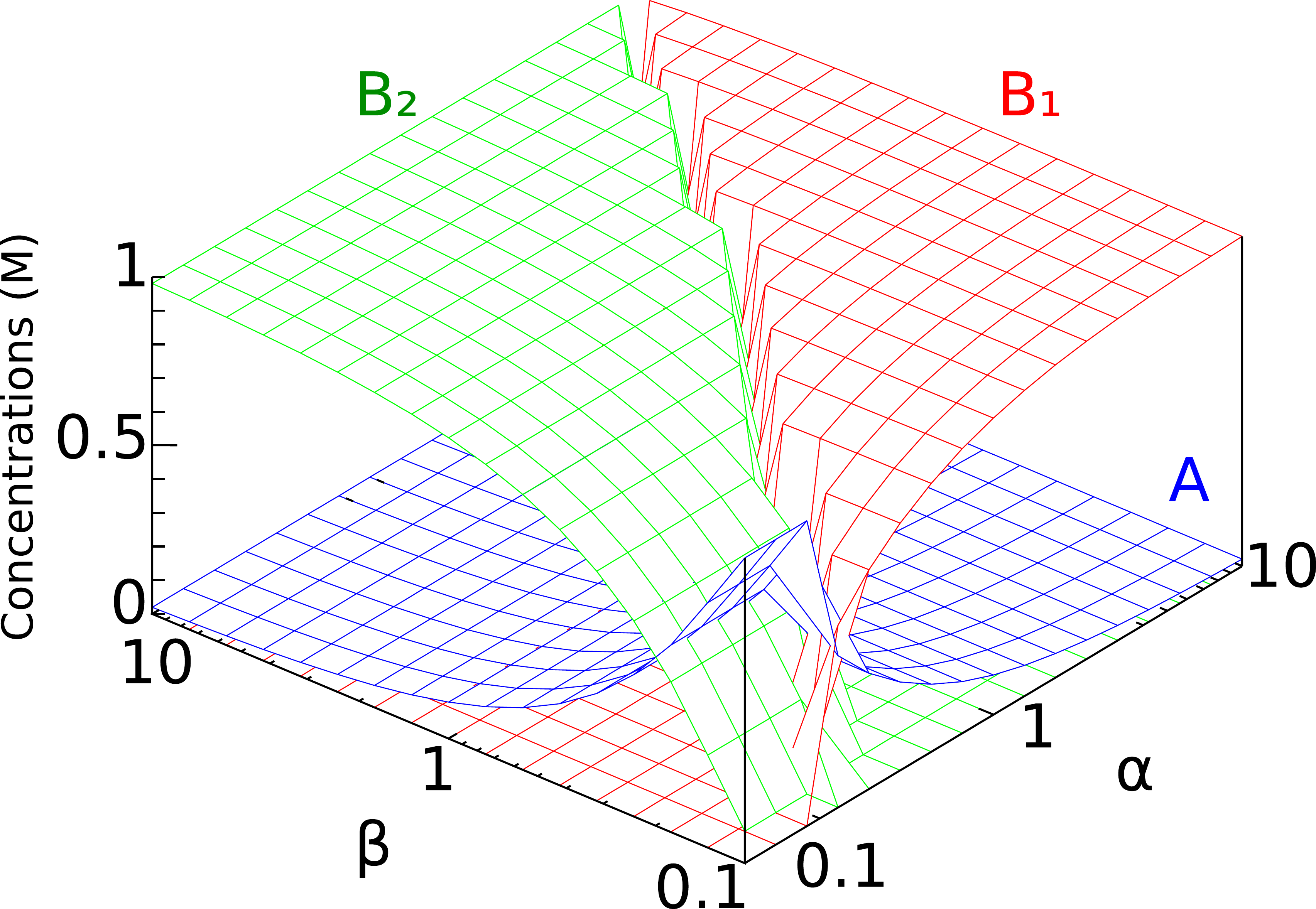}}

    \subfloat[Different zones of behaviors: majority of $A$ for
    $\alpha,\beta \ll 1$, majority of $B_1$ for $\alpha > \beta $,
    majority of $B_2$ for $\alpha < \beta $, and coexistence of $B_1$
    and $B_2$ for $\alpha,\beta \gg
    1$.]{\includegraphics[width=8cm]{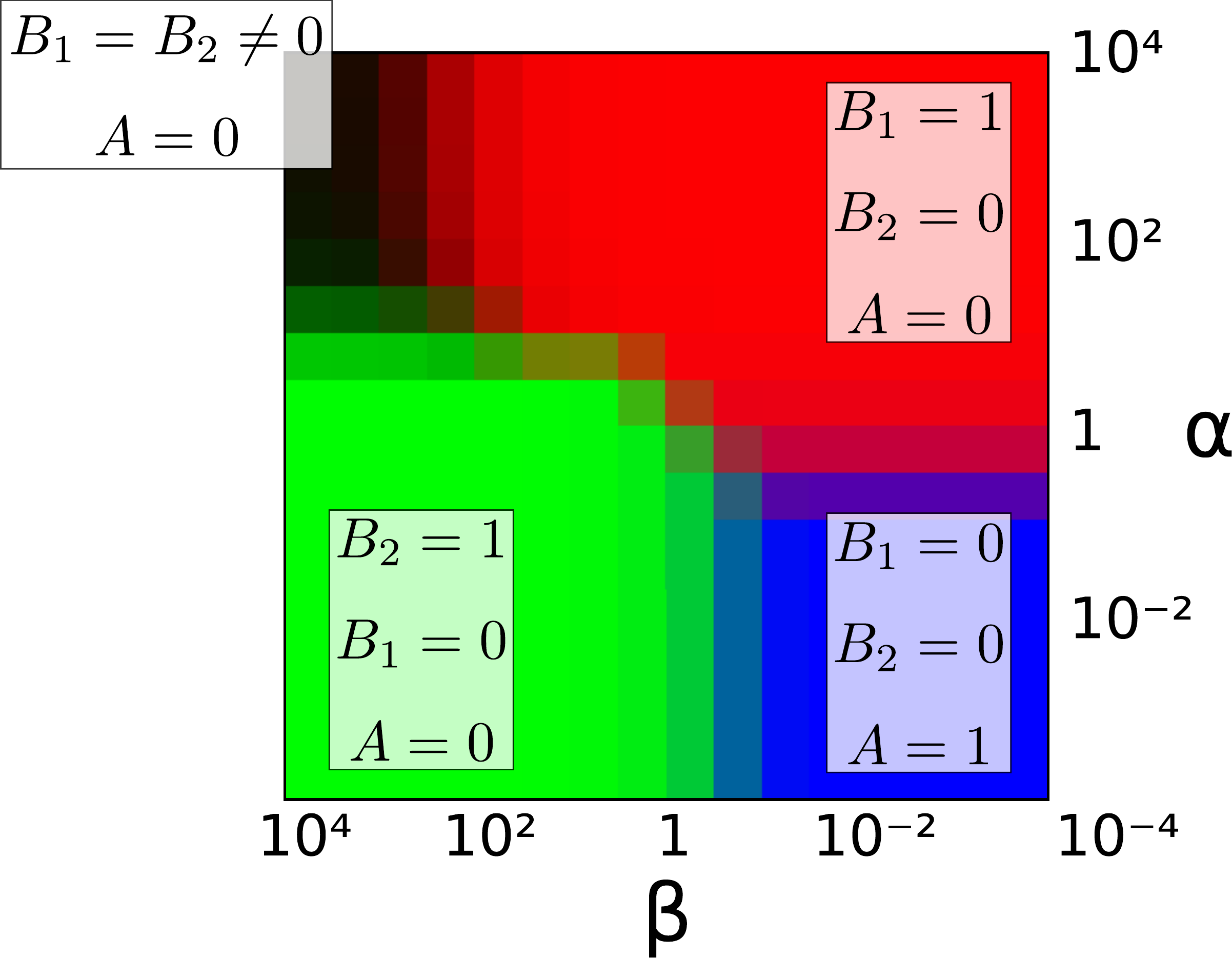}}

    \caption{Competition between template and autoinductive
      autocatalysis, generating respectively $B_1$ and $B_2$ compounds
      from the same $A$ compound. Incoming flux of $A$, and outgoing
      fluxes of $B_1$ and $B_2$, $10^{-5}$~M.s$^{-1}$. $K_A=1$,
      $K_{B_1}=K_{B_2}=100$. Direct autocatalysis:
      $\Gamma_{AC}=10^{-2}\cdot \alpha$, $\Gamma_{NC}=10^{-6}\cdot
      \alpha$. Autoinduction, according to \cref{fig1}(d):
      $\Gamma_{1}= \beta$, $\Gamma_{2}= \Gamma_{3}=100 \cdot \beta$,
      $K_C=K_E=1$; $K_{E^*}=10$.}
    \label{fig3}
  \end{center}
\end{figure}

The kinetic equivalence between template autocatalysis and
autoinductive autocatalysis can be shown by making these two
mechanisms to compete, replacing \cref{eq:mech1} and
\cref{eq:mech2} by the corresponding mechanism. Kinetic parameters
have first been normalized so that each reaction leads on their own to
the same kinetic behavior (sigmoidal evolution, half-reaction at
$10^5$~s), and then multiplied by respectively $\alpha$ and $\beta$
parameters in order to tune the respective velocity of each
mechanism. The result is actually symmetrical between the two
processes and only the fastest product is maintained in the system:
$B_1$ when $\alpha > \beta$, and $B_2$ when $\alpha < \beta$ (see
\cref{fig3}(a)). As a consequence,  while mechanistically different, these two
autocatalysis are shown to be dynamically equivalent.

This selectivity is independent of the relative stability of $B_1$ and
$B_2$, but is only possible for kinetics that are well adapted to the
global influx of matter. For slow kinetics, there is a flush of the
system, and neither $B_1$ nor $B_2$ can be maintained. For fast
kinetics, the system is close to equilibrium, the compounds $B_1$ and
$B_2$ being both present in proportion to their respective stability
(see \cref{fig3}(b)). Such result is well known for open flow Frank
systems \cite{Cruz.Parmananda.ea-08}.

\subsection{From Autocatalytic Processes towards Autocatalytic Sets}
\label{sec:autoc-proc-vs}

These competitive systems are able to dynamically maintain a set of components, to the detriment of
others. These autocatalytic networks must however not be confused with autocatalytic sets. This
latter notion is rather popular in the artificial life literature, but relies much more on the
cooperation between autocatalytic mechanisms than on the competition that has just been detailed
here. This implies a notion of material closure of the system and of self maintenance of the whole
network by crossing energetical fluxes\cite{Kauffman-86,Hordijk.Steel-04,Benko.Centler.ea-09}.
Confusion among these different phenomena can be pinpointed in the literature \cite{Blackmond-09},
when the failure of autoinductive sets to be maintained do not originate from a difference of
behavior between autocatalytic and autoinductive mechanisms, but from a defect in the closure of the
system (e.g. induced by the leakage of some components).

\section{Conclusion}
\label{sec:conclusion}

Important distinctions need to be made between mechanistic and dynamic aspects of autocatalysis. One
single mechanism can produce different dynamics, while identical dynamics can originate from
different mechanisms.  Thus, a pragmatic definition of autocatalysis have to be based on a kinetic
signature, in order to classify the systems according their observable behavior, rather than on a
mechanistic signature, that would instead classify the systems according to the origin of their
behavior.  All the different autocatalytic processes described in this work are able to generate
autocatalytic kinetics. They can constitute a pathway towards the onset of ``self-sustaining
autocatalytic sets'', as chemical attractor in non-equilibrium networks. However, the problem of
the evolvability of such systems must be kept in mind \cite{Vasas.Szathmary.ea-10}. If a system
evolves towards a stable attractor, no evolution turns out to be possible. There is the necessity of
``open-ended'' evolution \cite{Ruiz-Mirazo-07} i.e.\ the possibility for a dynamic set to not only
maintain itself (i.e.\ as a strict autocatalytic system) but also to act as a ``general
autocatalytic set'', redounding upon the concept originally introduced by Muller\cite{Muller-22} for the
autocatalytic power linked to mutability of genes. For example, insights can be gained by a deeper
and renewed study of the evolution of prions as a simple mechanism of mutable autocatalytic systems
\cite{Li.Browning.ea-10}.

\section{Appendix}
\label{sec:appendix}

The kinetic behavior of three different mechanisms for autocatalytic transformations have been
studied in details. The methodology consists in establishing the different chemical fluxes of the
network. The relationship between these fluxes can be simplified by assuming the QSSA for relevant
compounds. The purpose is then to establish the expression of the transformation flux $\varphi$ as a
function of the concentration of the reactants and the products.

\subsection{Indirect autocatalysis}
\label{sec:indir-autoc}

The four fluxes of \cref{fig1}(c) can be written as:
\begin{align}
  \varphi_1 &= \Gamma_1 (V_AV_D-V_C)  \label{eq:7}\\
  \varphi_2 &= \Gamma_2 (V_C-V_BV_E) \label{eq:9}\\
  \varphi_3 &= \Gamma_3 (V_E-V_B) \label{eq:10}\\
  \varphi_4 &= \Gamma_4 (V_B-V_D) \label{eq:11}
\end{align}
The QSSA for $D$ comes down to $\varphi_1\simeq\varphi_4$:
\begin{align}
  \Gamma_1 V_AV_D - \Gamma_1 V_C &= \Gamma_4 V_B - \Gamma_4 V_D   \label{eq:8}\\
(\Gamma_1 V_A + \Gamma_4) V_D &= \Gamma_4 V_B + \Gamma_1 V_C \label{eq:24}\\
V_D &=  \frac{\Gamma_4 V_B + \Gamma_1 V_C}{\Gamma_1 V_A + \Gamma_4} \label{eq:25}
\end{align}
Replacing $V_D$ by \cref{eq:25} in \cref{eq:7} gives:
\begin{align}
  \varphi_1 &= \Gamma_1
  \left(
    V_A \frac{\Gamma_4V_B + \Gamma_1 V_C}{\Gamma_1 V_A + \Gamma_4}-V_C)
  \right) \label{eq:13} \\
  &= \frac{\Gamma_1\Gamma_4}{\Gamma_1 V_A + \Gamma_4}V_AV_B +\Gamma_1V_C\frac{-\Gamma_4}{\Gamma_1
    V_A + \Gamma_4} \label{eq:14}\\
&= \frac{\Gamma_1\Gamma_4}{\Gamma_1 V_A + \Gamma_4} (V_AV_B-V_C) \label{eq:15}
\end{align}
The QSSA for $E$ comes down to $\varphi_2 \simeq \varphi_3$:
\begin{align}
\Gamma_2V_C - \Gamma_2 V_BV_E &= \Gamma_3 V_E - \Gamma_3V_B  \label{eq:16} \\
V_E &= \frac{\Gamma_2V_C+\Gamma_3V_B}{\Gamma_3+\Gamma_2V_B} \label{eq:17} 
\end{align}
Replacing $V_E$ by \cref{eq:17} in \cref{eq:9} by Eq.gives:
\begin{align}
  \varphi_2 &= \Gamma_2
  \left(
    V_C - V_B \frac{\Gamma_2V_C+\Gamma_3V_B}{\Gamma_3+\Gamma_2V_B}
  \right) \label{eq:18}\\
  &= \Gamma_2 V_C \frac{\Gamma_3}{\Gamma_3 + \Gamma_2V_B} -\Gamma_2V_B \frac{\Gamma_3 V_B}{\Gamma_3
  + \Gamma_2V_B}\label{eq:19}\\
&=\frac{\Gamma_2 \Gamma_3}{\Gamma_3+\Gamma_2V_B}(V_C-V_B^2) \label{eq:20}
\end{align}
At last, the QSSA for $C$ comes down to $\varphi_1\simeq\varphi_2=\varphi$. Combining \cref{eq:15} and
Eq,~\cref{eq:20} gives:
\begin{align}
  V_C &= \frac{\Gamma'_1 V_AV_B + \Gamma'_2 V_B^2}{\Gamma'_1 + \Gamma'_2}\label{eq:21} \\
\mathrm{with}\qquad \Gamma'_1 &=\frac{\Gamma_1\Gamma_4}{\Gamma_1 V_A + \Gamma_4}  \label{eq:22} \\
\mathrm{and}\qquad \Gamma'_2 &= \frac{\Gamma_2 \Gamma_3}{\Gamma_3+\Gamma_2V_B} \label{eq:23}
\end{align}
Replacing $V_C$ by \cref{eq:21} in \cref{eq:15} gives:
\begin{align}
  \varphi &= \Gamma'_1 V_AV_B - \Gamma'_1  \frac{\Gamma'_1 V_AV_B + \Gamma'_2 V_B^2}{\Gamma'_1 +
    \Gamma'_2}\label{eq:26}\\
  &=
  \left(
    \Gamma'_1 -\frac{\Gamma'^2_1}{\Gamma'_1+\Gamma'_2}
  \right)V_AV_B
  - \frac{\Gamma'_1\Gamma'_2}{\Gamma'_1+\Gamma'_2}V_B^2 \label{eq:27}\\
  &= \frac{\Gamma'_1\Gamma'_2}{\Gamma'_1 + \Gamma'_2} (V_AV_B-V_B^2) \label{eq:28}
\end{align}
Replacing $\Gamma'_1$ and $\Gamma'_2$ by their expression given in \cref{eq:22} and \cref{eq:23}
then gives:
\begin{align}
\varphi &= \frac{V_AV_B - V_B^2}{\frac{1}{\Gamma_1}+\frac{1}{\Gamma_2}+\frac{V_A}{\Gamma_4}+\frac{V_B}{\Gamma_3}}  \label{eq:29}
\end{align}

\subsection{Autoinductive autocatalysis}
\label{sec:auto-autoc}

The three fluxes of \cref{fig1}(d) are:
\begin{align}
  \varphi_1 &= \Gamma_1(V_AV_E-V_C)\label{eq:12}\\
  \varphi_2&= \Gamma_2(V_C - V_BV_E) \label{eq:30}\\
  \varphi_3 &= \Gamma_3(V_BV_{E^*} - V_C) \label{eq:31}
\end{align}
The QSSA for $C$ comes down to $\varphi_1+\varphi_3 \simeq\varphi_2$, and the QSSA for $E$ comes down to
$\varphi_1 \simeq \varphi_2$. This implies that $\varphi_3 \ll \varphi_1$, so that with
$\varphi_3=\varepsilon$ and, $\varphi_1 = \varphi$, we obtain:
\begin{align}
\varphi_2&=\varphi+\varepsilon\simeq \varphi\label{eq:36}
\end{align}
In that context, \cref{eq:31} gives:
\begin{align}
  V_C &= V_BV_{E^*}-\frac{\varepsilon}{\Gamma_3}\label{eq:34}
\end{align}
Combining \cref{eq:12}, \cref{eq:30} in \cref{eq:36} then gives: 
\begin{align}
  \Gamma_2 V_C -\Gamma_2V_BV_E  &= \Gamma_1V_AV_E -\Gamma_1V_C + \varepsilon\label{eq:35}\\
   V_E &= \frac{(\Gamma_1+\Gamma_2)V_C - \varepsilon}{\Gamma_1V_A + \Gamma_2 V_B} \label{eq:33}
\end{align}
Replacing $V_C$ by its value given in \cref{eq:34} leads to:
\begin{align}
  V_E &= \frac{(\Gamma_1+\Gamma_2)V_{E^*}V_B - \frac{\Gamma_1 + \Gamma_2 +
      \Gamma_3}{\Gamma_3}\varepsilon}{\Gamma_1V_A + \Gamma_2 V_B} \label{eq:37}\\
  V_E &\simeq \frac{(\Gamma_1+\Gamma_2)V_{E^*}V_B}{\Gamma_1V_A + \Gamma_2 V_B} \label{eq:38}
\end{align}
The flux of destruction of $A$ can be computed by replacing $V_E$ in \cref{eq:12} by
\cref{eq:38} (computing the flux of formation of $B$ from \cref{eq:30} would of course
give the same result):
\begin{align}
 \varphi &= \Gamma_1
 \left(
   V_A\frac{(\Gamma_1+\Gamma_2)V_{E^*}V_B}{\Gamma_1V_A + \Gamma_2 V_B} - V_BV_{E^*}
 \right)\label{eq:39}\\
 &= \Gamma_1
   \frac{(\Gamma_1+\Gamma_2)V_{E^*}V_BV_A - V_BV_{E^*}(\Gamma_1V_A + \Gamma_2 V_B)}{\Gamma_1V_A +
     \Gamma_2 V_B} \label{eq:40} \\
 &= \Gamma_1\Gamma_2V_{E^*}
   \frac{V_BV_A - V_B^2}{\Gamma_1V_A +
     \Gamma_2 V_B} \label{eq:41}
\end{align}
The law of conservation of $E$ compounds leads to:
\begin{align}
  \varphi&=
   \frac{ \Gamma_1\Gamma_2V_{E^*}^0( V_BV_A - V_B^2)}{(\Gamma_1V_A +
     \Gamma_2 V_B)(1+r_CV_C+r_EV_E)} 
   \label{eq:43}
\end{align}
with $r_C=K_C/K_{E^*}$ and $r_E=K_E/K_{E^*}$. Assuming that $E^*$ is the much more stable than $C$
and $E$, $r_C$ and $r_E \ll 1$, so that we finally obtain\footnote{Without the hypothesis of a large
  stability of $E^*$, not neglecting the $r_C$ and $r_E$ terms eventually leads to add $V_B$ terms to the
  denominator, which will tend to destroy the autocatalytic effect.}:
\begin{align}
  \varphi&=
   \frac{ \Gamma_1\Gamma_2V_{E^*}^0}{\Gamma_1V_A +
     \Gamma_2 V_B} 
   (  V_BV_A - V_B^2)\label{eq:42}
\end{align}

\subsection{Iwamura's model}
\label{sec:iwamuras-model}

The five fluxes of \cref{fig1}(e) are:
\begin{align}
  \varphi_1 &= \Gamma_1(V_AV_P-V_{PA})\label{eq:53}\\
  \varphi_2&=\Gamma_2(V_{PA}V_X-V_{PAX})\label{eq:44}\\
  \varphi_3&=\Gamma_3(V_{PAX}-V_{P|AX})\label{eq:45}\\
  \varphi_4&=\Gamma_4(V_{P|AX}-V_PV_{AX})\label{eq:46}\\
  \varphi_5&=\Gamma_5(V_{P|AX}V_A-V_{PA}V_{AX})\label{eq:47}
\end{align}
The QSSA for $P$ leads to $\varphi_1=\varphi_4$; for $PA$, it leads to
$\varphi_2=\varphi_1+\varphi5$; for $PAX$ it leads to $\varphi_3=\varphi_2$; for $P|AX$, it leads to
$\varphi_3=\varphi_4+\varphi_5$. The fluxes can thus be decomposed into two elementary fluxes:
\begin{align}
  \varphi_1 &= \varepsilon \label{eq:48}\\
  \varphi_2 &= \varphi + \varepsilon \label{eq:49} \\
  \varphi_3 &= \varphi + \varepsilon \label{eq:50}\\
  \varphi_4 &= \varepsilon \label{eq:51} \\
  \varphi_5 &= \varphi \label{eq:52}
\end{align}
$\varphi$ is the flux of the catalytic reaction, and $\varepsilon$ the flux of the non-catalytic
reaction, so that $\varepsilon \ll \varphi$. This would typically be characterized by $\Gamma_1 \ll
\Gamma_5$. 

$\varphi_2=\varphi_3$ leads to:
\begin{align}
V_{P|AX} &= \frac{\Gamma_{2}}{\Gamma_{23}}V_{PAX}-\frac{\Gamma_2}{\Gamma_3}V_{PA}V_{X}  \label{eq:32}
\end{align}
with $\Gamma_{23}=\Gamma_2\Gamma_3/(\Gamma_2+\Gamma_3)$

$\varphi_3 \simeq \varphi_5$ leads to:
\begin{align}
V_{P|AX} &= \frac{\Gamma_{23}}{\Gamma_{2}}V_{PAX}+\frac{\Gamma_5}{\Gamma_3+\Gamma_5V_A}V_{PA}V_{AX}  \label{eq:55}
\end{align}
Combining \cref{eq:32} and \cref{eq:55}, eliminating $V_{P|AX}$ leads to:
\begin{align}
 V_{PAX}&=\frac{\frac{\Gamma_5}{\Gamma_2}V_{AX}+V_X+\frac{\Gamma_5}{\Gamma_3}V_AV_X}{\Gamma_5 V_A +
 \Gamma_{23}}\Gamma_{23}V_{PA} \label{eq:54}
\end{align}
Combining \cref{eq:54} with \cref{eq:32} leads to:
\begin{align}
 V_{P|AX}&=\frac{\Gamma_5V_{AX}+\Gamma_{23}V_X}{\Gamma_5V_A+\Gamma_{23}}V_{PA}  \label{eq:56}
\end{align}

$\varphi_1 = \varphi_4$ leads to:
\begin{align}
 V_P&=\frac{\Gamma_4V_{P|AX}+\Gamma_1V_{PA}}{\Gamma_1 V_A+\Gamma_4V_{AX}} \label{eq:57}
\end{align}
Combining \cref{eq:57} and \cref{eq:56} leads to:
\begin{equation}
  \label{eq:58}
  V_P=\frac{\Gamma_4(\Gamma_5V_{AX}+\Gamma_{23}V_X) + \Gamma_1(\Gamma_5V_A+\Gamma_{23})}
  {(\Gamma_1V_A+\Gamma_4 V_{AX})(\Gamma_5V_A+\Gamma_{23})}V_{PA}  
\end{equation}
The flux of production of AX can be computed from \cref{eq:44}, \cref{eq:45} or \cref{eq:47},
which leads to:
\begin{align}
  \label{eq:59}
  \varphi&=\Gamma_5\Gamma_{23}V_{PA}\frac{V_AV_X-V_{AX}}{\Gamma_5V_A+\Gamma_{23}}
\end{align}
Combining \cref{eq:59} and \cref{eq:58} leads to:
\begin{equation}
  \label{eq:60}
  \varphi = 
\frac
{\Gamma_5\Gamma_{23}V_P(V_AV_X-V_{AX})
  \left(
    \frac{\Gamma_1}{\Gamma_4}V_A+V_{AX}
  \right)}
{\Gamma_5
  \left(
    \frac{\Gamma_1}{\Gamma_4}V_A+V_{AX}
  \right)+\Gamma_{23}
  \left(
    \frac{\Gamma_1}{\Gamma_4}+V_X
  \right)}
\end{equation}
This can be simplified in an ideal case, assuming that the compound $P$ is the mores stable compound
among $P$, $PA$, $PAX$ and $P|AX$, so that $V_P \simeq V_P^0$, and that the reactivities are so that
$\Gamma_1 \ll (\Gamma_4, \Gamma_5) \ll \Gamma_{23}$ (i.e.\ assuming that reaction 1 is very slow, and
that reactions 2 and 3 are very fast), which leads to:
\begin{equation}
  \label{eq:61}
  \varphi \simeq \Gamma_5 V_P^0
  \left(
    V_AV_{AX}-\frac{V_{AX}^2}{V_X}
  \right)
\end{equation}

\small

\paragraph*{Acknowledgment:}
This work was done within the scope of the European program COST
 ``System Chemistry'' CM0703. We additionally thank R. Pascal for
 useful discussions.

\bibliographystyle{naturemag}
\bibliography{biblio}

\balance

\end{document}